\documentclass[showpacs,showkeywords,aps,graphicx,twocolumn]{revtex4}

\usepackage{graphicx}

\usepackage{array}

\begin{document}

\title{Device-independent quantum secure direct communication with single photon sources }

\author{Lan Zhou$^{1}$, Bao-Wen Xu$^{1,2}$, Wei Zhong$^{3}$, Yu-Bo Sheng $^{2,3}$\footnote{Email address:
shengyb@njupt.edu.cn} }
\address{$^1$College of Science, Nanjing University of Posts and Telecommunications, Nanjing,
210023, China\\
$^2$College of Electronic and Optical Engineering, \& College of Flexible Electronics (Future Technology), Nanjing
University of Posts and Telecommunications, Nanjing, 210023, China\\
 $^3$Institute of Quantum Information and Technology, Nanjing University of Posts and Telecommunications, Nanjing, 210003, China\\}
\date{\today }

\begin{abstract}
Quantum secure direct communication (QSDC) can directly transmit secrete messages through quantum channel. Device-independent (DI) QSDC can guarantee the communication security relying only on the observation of the Bell inequality violation, but not on any detailed description or trust of the inner workings of users' devices. In the paper, we propose a DI-QSDC protocol with practical high-efficient single photon sources. The communication parties construct the entanglement channel from single photons by adopting the heralded architecture, which makes the message leakage rate independent of the photon transmission loss. The secure communication distance and the practical communication efficiency of the current DI-QSDC protocol are about 6 times and 600 times of those in the original DI-QSDC protocol. Combining with the entanglement purification, the parties can construct the nearly perfect entanglement channel and completely eliminate the message leakage. This DI-QSDC protocol may have important application in future quantum communication field.

\textbf{Keywords:} device-independent quantum secure direct communication, single-photon source, Bell state measurement, secret message capacity, practical communication efficiency
\end{abstract}
\pacs{ 03.67.Pp, 03.67.Hk, 03.65.Ud} \maketitle

\section{Introduction}
Quantum secure communication, based on intrinsic
properties of quantum systems, can guarantee the absolute security of communication. Quantum key distribution and quantum secure direct communication (QSDC) are two important branches of quantum secure communication. Quantum key distribution can distribute secure keys between the sender and the receiver, which was proposed in 1984 \cite{BB84}. Quantum key distribution has been widely investigated
in both theory and experiment \cite{QKD1,addqkd4,QKD2,QKD3,QKD4,QKD7,qkd7nn,qkd7n,QKD8,QKD17n,qkd9,qkd10,qkd11,qkd12,qkd13,qkd14,qkd14n,qkd15,qkd15n,QKD15n}. QSDC allows the message sender to directly transmit secret messages to the receiver without keys \cite{qsdc1,qsdc2,qsdc3,qsdc4n,qsdc4,qsdc5,qsdc6,qsdc8,qsdc9,qsdc10,addqsdc1,addqsdc2,addqsdc3,qsdc11,qsdc12,qsdc12n,qsdc13,qsdc14,qsdc15,qsdc14n,qsdc15n,qsdc16n,qsdc16,qsdc17,qsdc18,qsdc19}. QSDC protocol was proposed in 2000 \cite{qsdc1}. Later, the typical entanglement-based two-step QSDC protocol and single-photon-based QSDC protocol were successively proposed \cite{qsdc2,qsdc3}, which were experimentally demonstrated in 2016 and 2017, respectively \cite{qsdc4,qsdc5}. In 2020, the device-independent (DI) QSDC and measurement-device-independent QSDC protocols were put forward, which can guarantee QSDC's security under practical experimental condition \cite{qsdc9,qsdc10}. In 2022, the one-step QSDC was put forward, which can simplify the operation and reduce the message loss \cite{qsdc16}.
During past few years, QSDC has made great experimental
progress. In 2021, a 15-user QSDC network with any two users being 40 km apart was demonstrated \cite{qsdc14}. Recently, researchers achieved the QSDC over 100 km fiber with time-bin and phase quantum states \cite{qsdc15}.

Similar as DI quantum key distribution \cite{DIQKD2,DIQKD3,DIQKD4,DIQKD6,DIQKD7,DIQKD8,DIQKD9,DIQKD10}, DI-QSDC relaxes conventional assumptions on devices and allows users to transmit secrete messages with unknown and uncharacterized devices. As long as some minimal
assumptions (the quantum physics is correct and no unwanted signal can escape from the communication parties' laboratories) are satisfied, DI-QSDC can guarantee the communication security based solely on
the observed data conclusively violating the Bell inequality (typically, the Clauser-Horne-Shimony-Holt (CHSH) inequality) \cite{Bell1,Bell2,CHSH1}. The observation of the CHSH inequality violation should close the so-called detection loophole \cite{Bell1}. Although recent advances on single photon detector have achieved the detection efficiency close to 1 \cite{detector1,detector2}, DI-QSDC still faces big challenges. On one hand, the original DI-QSDC protocol requires the entanglement photon source and constructs the entanglement channel by the long-distance entanglement distribution. The entanglement generation of practical entanglement source (spontaneous parametric down-conversion source) is probabilistic and the double-pair emission can not be eliminated \cite{spdc,spdc1,hyper6}. On the other hand, the experimental devices and quantum channel are imperfect, which may cause photon loss. Photon loss occurring at the
generation, transmission, and detection stages would deteriorate the nonlocal
correlations between the photons. Above two features provide an opportunity for the eavesdropper (Eve) to steal some photons without being detected and largely reduce the secure communication distance.

  Actually, comparing with the spontaneous parametric down-conversion source, the practical single photon sources have already
allowed for nearly on-demand \cite{singlephoton1}, highly-efficient \cite{singlephoton2} extraction of single photons (also in pulse
trains \cite{singlephoton3,singlephoton4} as well as at telecom wavelengths \cite{singlephoton5}). Current single photon sources can maintain the purity and indistinguishability of the generated photons with the probability of above 99\% \cite{singlephoton6,singlephoton7}. In 2020, Long \emph{et at.} reported an experimental implementation of free-space QSDC based on single
photon sources with the repetition rate of 16 MHz \cite{addqsdc2}. In 2022, a high-fidelity photonic quantum logic gate
based on near-optimal Rydberg single-photon source was realized. The excitation frequencies of the 780 nm and 479 nm laser pulses reach $\Omega^{e}_{780}/2\pi\approx 6.4$ MHz and $\Omega^{e}_{479}/2\pi\approx 4.2$ MHz, respectively \cite{source7}. Soon later, a single-photon source at the telecom wavelength based on InAs/GaAs quantum dots was reported. This single photon source has the bright single-photon emission with Purcell factor $>5$
and count rates up to 10 MHz \cite{source8}. Meanwhile, the heralded architectures have been used to construct the entanglement channel, which can eliminate the influence from photon transmission loss on the Bell (CHSH) violations \cite{DIQKD4}. In 2020,
two DI quantum key distribution schemes were proposed, which realized the entanglement creation process
with single-photon sources \cite{DIQKD10}, and distributed keys
at high rates over large distance.
In this work, we propose a DI-QSDC protocol based on single photon sources and the heralded complete Bell state measurement (CBSM). Comparing with original DI-QSDC protocol, this current DI-QSDC protocol can efficiently increase the practical communication efficiency and communication distance.

The paper is organized as follows. In Sec. II, we introduce the heralded long-distance entanglement distribution based on single-photon sources. In Sec. III, we explain the DI-QSDC protocol in detail. In Sec. IV, we provide the security analysis and calculate the secrecy message capacity and practical communication efficiency of the DI-QSDC protocol against collective attacks. In Sec. V, we make a discussion and finally provide a conclusion.

\section{Heralded long-distance entanglement distribution based on single photon sources}
\begin{figure}[!h]
\begin{center}
\includegraphics[width=9cm,angle=0]{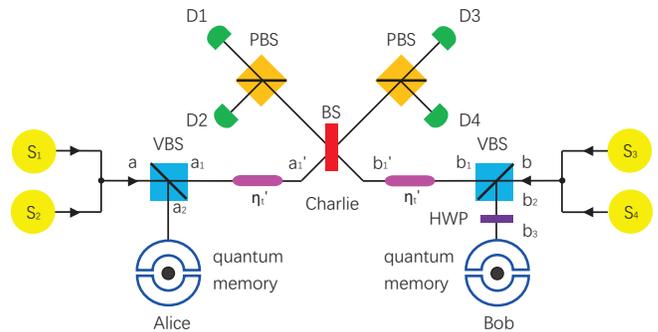}
\caption{The basic principle of constructing the long-distance entanglement channel between Alice and Bob with the help of the \textbf{heralded architecture}. Here, $S_{i}$ ($i=1,2,3,4$) represents the single photon source. The generated photons pass through a variable beam splitter (VBS). After the VBS, the photons in the transmitted port of the VBSs are sent to Charlie for the Bell state measurement (BSM). $\eta_{t}'$ represents the photon transmission efficiency corresponding to the distance $L_{A(B)C}$ between Alice (Bob) and Charlie. BS and PBS represent the 50:50 beam splitter and polarization beam splitter, respectively. HWP represents the half wave plate.}
\end{center}
\end{figure}

We  explain the construction of the long-distance entanglement channel based on the single photon sources \cite{entanglementpra}. From Ref. \cite{singlephoton1,singlephoton2,singlephoton3,singlephoton4,singlephoton5,singlephoton6,singlephoton7,source7,source8}, we can treat the single photon sources as the on-demand sources.
In contrast, the practical spontaneous parametric down-conversion source adopted in the original DI-QSDC protocol \cite{qsdc9} can generate the mixed state as \cite{spdc}
\begin{eqnarray}
\rho_{0}\approx(1-p-p^{2})|vac\rangle\langle vac|+p|\phi\rangle\langle\phi|+p^{2}|\phi^{\otimes2}\rangle\langle\phi^{\otimes2}|,\label{spdc}
\end{eqnarray}
where $|vac\rangle$ means the vacuum state, $|\phi\rangle=\frac{1}{\sqrt{2}}(|HH\rangle+|VV\rangle)$. The desired entangled photon pair generates at the  order of $p$ ($p\sim10^{-5}-10^{-3}$) \cite{hyper6}. Meanwhile, the double pair emission is unavoidable.

As shown in Fig. 1, Alice (Bob) adopts two single photon sources  $S_{1}$ and $S_{2}$ ($S_{3}$ and $S_{4}$) to prepare two single photons with the horizontal polarization ($|H\rangle$) and the vertical polarization ($|V\rangle$), respectively.

Then, each of the two parties passes his photons through a variable beam splitter (VBS) with the transmittance of $T$, which makes the photon states in Alice's and Bob's locations evolve to
\begin{eqnarray}
|\phi_{1}\rangle&=&|H\rangle_a\otimes|V\rangle_a\nonumber\\
&\rightarrow&(\sqrt{T}|H\rangle_{a_{1}}+\sqrt{1-T}|H\rangle_{a_{2}})\nonumber\\
&\otimes&(\sqrt{T}|V\rangle_{a_{1}}+\sqrt{1-T}|V\rangle_{a_{2}}),\nonumber\\
|\phi_{2}\rangle&=&|H\rangle_b\otimes|V\rangle_b\nonumber\\
&\rightarrow&(\sqrt{T}|H\rangle_{b_{1}}+\sqrt{1-T}|H\rangle_{b_{2}})\nonumber\\
&\otimes&(\sqrt{T}|V\rangle_{b_{1}}+\sqrt{1-T}|V\rangle_{b_{2}}).
\end{eqnarray}
Alice and Bob send the photon in $a_{1}$ and $b_{1}$ modes to a third party Charlie for the Bell state measurement (BSM). The BSM devices are totally in linear optics, which can only distinguish $|\psi^{\pm}\rangle_{a_{1}'b_{1}'}$.
In detail, a click in $D1D2$, or $D3D4$ indicates a projection into $|\psi^{+}\rangle_{a_{1}'b_{1}'}$, and a click in $D2D3$, or $D1D4$  projects the quantum state into $|\psi^{-}\rangle_{a_{1}'b_{1}'}$ \cite{entanglementpra}. Then, Bob passes the reflected photon in $b_{2}$ mode through the half wave plate. Next, the photon in $a_{2}$ and $b_{3}$ modes are stored in the quantum memory devices. The experimental realizations of the quantum memory in the single-photon level with the electromagnetically induced transparency have been reported since 2013 \cite{memory,memory1}. It is noticeable that the quantum memory can herald the existence of the photon in the reflected port. Only when the BSM is successful and the quantum memory in each party's location responses, the distant parties can share the entangled state. Otherwise, the entanglement distribution will fail. In this way, only the case that one photon transmits the VBS and the other photon is reflected by the VBS in each party's location may lead to the success of the entanglement distribution. We define the photon transmission efficiency $\eta_{t}'=10^{-\alpha L_{A(B)C}/10}$ \cite{loss} corresponding to the distance $L_{A(B)C}$ between Alice (Bob) and Charlie, where $\alpha=0.2$ $dB/km$.
If the transmitted photons lose during the transmission process, the BSM cannot obtain the successful detector response. As a result, for obtaining the successful entanglement distribution, the state $|\phi_{1}\rangle\otimes|\phi_{2}\rangle$ will collapse to
\begin{eqnarray}
&&|\Phi_{1}\rangle=\sqrt{\eta_{t}'T(1-T)}(|HV\rangle_{a_{1}'a_{2}}+|VH\rangle_{a_{1}'a_{2}})\nonumber\\
&\otimes&\sqrt{\eta_{t}'T(1-T)}(|HV\rangle_{b_{1}'b_{2}}+|VH\rangle_{b_{1}'b_{2}})\nonumber\\
&=&\eta_{t}'T(1-T)(|HH\rangle_{a_{1}'b_{1}'}|VV\rangle_{a_{2}b_{2}}+|VH\rangle_{a_{1}'b_{1}'}|HV\rangle_{a_{2}b_{2}}\nonumber\\
&+&|HV\rangle_{a_{1}'b_{1}'}|VH\rangle_{a_{2}b_{2}}+|VV\rangle_{a_{1}'b_{1}'}|HH\rangle_{a_{2}b_{2}})\nonumber\\
&=&\frac{\eta_{t}'T(1-T)}{\sqrt{2}}(|\phi^{+}\rangle_{a_{1}'b_{1}'}|\phi^{+}\rangle_{a_{2}b_{2}}-|\phi^{-}\rangle_{a_{1}'b_{1}'}|\phi^{-}\rangle_{a_{2}b_{2}}\nonumber\\
&+&|\psi^{+}\rangle_{a_{1}'b_{1}'}|\psi^{+}\rangle_{a_{2}b_{2}}-|\psi^{-}\rangle_{a_{1}'b_{1}'}|\psi^{-}\rangle_{a_{2}b_{2}}),\label{bsm}
\end{eqnarray}
where $|\phi^{\pm}\rangle$ and $|\psi^{\pm}\rangle$ represent the polarization Bell states with the form of
\begin{eqnarray}
|\phi^{\pm}\rangle&=&\frac{1}{\sqrt{2}}(|HH\rangle\pm|VV\rangle),\nonumber\\
|\psi^{\pm}\rangle&=&\frac{1}{\sqrt{2}}(|HV\rangle\pm|VH\rangle).\label{bell}
\end{eqnarray}
We define the total photon transmission efficiency $\eta_{t}=\eta_{t}'^{2}=10^{-0.2L_{AB}/10}$ for simplicity.

 As shown in Eq. (\ref{bsm}), if the BSM result is $|\psi^{+}\rangle_{a_{1}'b_{1}'}$, $|\Phi_{1}\rangle$ will collapse to $|\psi^{+}\rangle_{a_{2}b_{2}}$, which can be transformed to $|\phi^{+}\rangle_{a_{2}b_{3}}$ after the photon in $b_{2}$ passing through the half wave plate. If the BSM result is $|\psi^{-}\rangle_{a_{1}'b_{1}'}$, $|\Phi_{1}\rangle$ will collapse to $|\psi^{-}\rangle_{a_{2}b_{2}}$, which can be transformed  to  $|\phi^{-}\rangle_{a_{2}b_{3}}$ after the half wave plate. $|\phi^{-}\rangle_{a_{2}b_{3}}$ can  be further transformed to $|\phi^{+}\rangle_{a_{2}b_{3}}$ after the phase-flip operation. As a result, when the BSM is successful and the quantum memory in each party's location responses, Alice and Bob can deterministically obtain the pure output quantum state $|\phi^{+}\rangle_{a_{2}b_{3}}$ and the photon transmission loss case can be automatically eliminated. The probability of obtaining the successful entanglement distribution can be calculated as
\begin{eqnarray}
P_{1}=\eta_{t}'^{2}T^{2}(1-T)^{2}=\eta_{t}T^{2}(1-T)^{2}.\label{P1}
\end{eqnarray}
It can be easily found that $P_{1}$ can reach the maximal value of $\frac{\eta_{t}}{8}$ when $T=0.5$.

\section{The DI-QSDC protocol with single photon sources}
The security of the current DI-QSDC protocol can be guaranteed by only two fundamental assumptions. First, the quantum
physics is correct and Eve obeys the rules of quantum physics. Second, Alice's and Bob's physical locations are secure, say, no unwanted information can leak to
the outside. The basic principle of the DI-QSDC protocol is shown in Fig. 2.

\begin{figure*}
\centering
\includegraphics[width=14cm,angle=0]{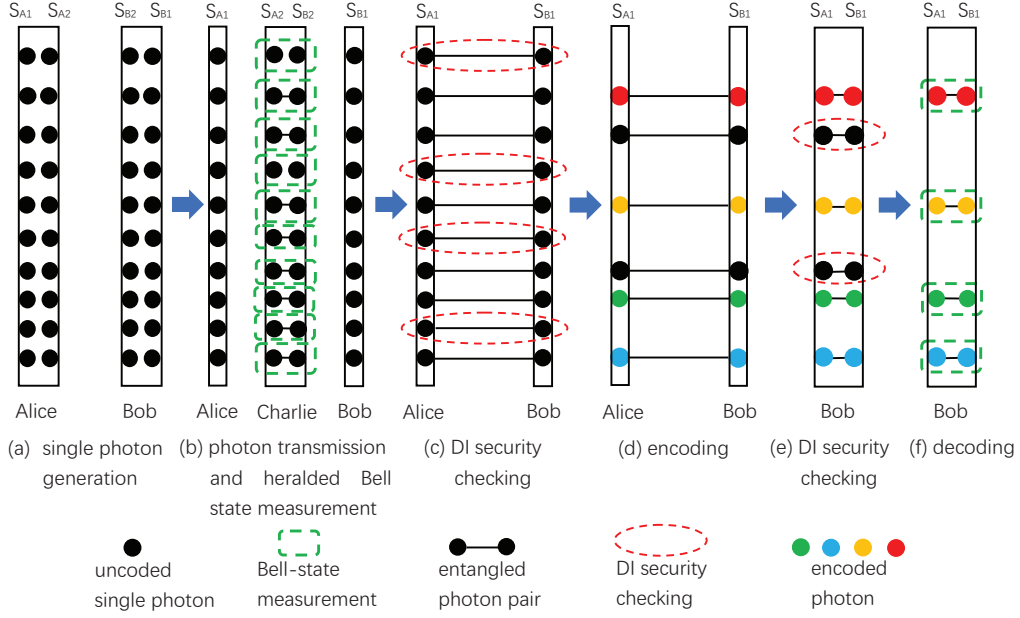}
\caption{Schematic principle of the DI-QSDC protocol. In the protocol, Alice and Bob prepare single photon sequences with orthogonal polarizations $|H\rangle$ and $|V\rangle$, respectively. They construct the entanglement channels by the heralded architecture. Then, Alice encodes the messages by two unitary operations and sends the encoded photons to Bob. After the photon transmission, Bob performs the BSM on each photon pair and can finally read out the encoded secure messages. The security of both photon transmission processes are guaranteed by the DI security checking.}
\end{figure*}

Step 1: Alice and Bob construct the entanglement channel based on the principle in Sec. II.  In detail, Alice and Bob each prepare ordered N (N is large) photon pairs in the state of $|H\rangle\otimes|V\rangle$ from the on-demand single-photon sources.
 Each of they passes the photons through a VBS. The photons in the reflected ports construct the $S_{A1}$ and $S_{B1}$ sequences, and those in the transmitted ports construct the $S_{A2}$ and $S_{B2}$ sequences. As shown in Fig. 1, Alice and Bob send the photon in $S_{A2}$ and $S_{B2}$ sequences to Charlie for the BSM and store the photons in $S_{A1}$ and $S_{B1}$ sequences in the quantum memory devices. Only when the BSM is successful and each quantum memory responses, the photons in the  $S_{A1}$ and $S_{B1}$ sequences can finally evolve to $|\phi^{+}\rangle$. Otherwise, Alice and Bob discard the photons in $S_{A1}$ and $S_{B1}$ sequences. In this way, Alice and Bob can deterministically construct the entanglement channel in $|\phi^{+}\rangle$ with $N_{1}$ entangled photon pairs, where $N_{1}=P_{1}N$ in theory.

Step 2: To ensure the security of the photon transmission, Alice randomly selects some photons in $S_{A1}$ sequence as the security checking photons and announces their positions to Bob through a public channel. They extract the security checking photons from the quantum memories to make the  round of DI security checking. In detail, for each security checking photon, Alice has four possible measurement bases  $A_{0}=\sigma_{z}$, $A_{1}=\frac{\sigma_{z}+\sigma_{x}}{\sqrt{2}}$, $A_{2}=\frac{\sigma_{z}-\sigma_{x}}{\sqrt{2}}$, and $A_{3}=\sigma_{x}$, and Bob has two possible measurement bases
$B_{1}=A_{0}$ and $B_{2}=A_{3}$ \cite{DIQKD8,DIQKD9}. All the measurement results $a=\{a_{0}, a_{1}, a_{2}, a_{3}\}$ and $b=\{b_{1}, b_{2}\}$ have binary outcome "$+1$" or "$-1$". Without loss of generality, we suppose that the marginal of all the measurements are random, such as $\langle a_{i}\rangle=\langle b_{j}\rangle=0$ ($i\in\{0,1,2,3\}$, $j\in\{1,2\}$). If the parties obtain the inconclusive result (the photon detectors click no photon), the measurement
result is set to be "+1" or "-1" randomly. After all the checking photon pairs have been measured, Alice and Bob announce their measurement bases and measurement results.

 There are four different cases.
In the  case, if Alice chooses $A_{1}$ or $A_{2}$ basis, their measurement results are used to estimate the CHSH polynomial as
 \begin{eqnarray}
S_{1}=\langle a_{1}b_{1}\rangle+\langle a_{1}b_{2}\rangle+\langle a_{2}b_{1}\rangle-\langle a_{2}b_{2}\rangle, \label{S}
\end{eqnarray}
where $\langle a_{i}b_{j}\rangle$ is defined as $P(a_{i}=b_{j})-P(a_{i}\neq b_{j})$ (the probability of $a_{i}=b_{j}$ minus the probability of $a_{i}\neq b_{j}$ ).
In the second case, if Alice chooses $A_{0}$ and Bob chooses $B_{1}$, their measurement results are used to estimate the quantum bit-flip error rate $Q_{b1}$ as
\begin{eqnarray}
Q_{b1}=P(a_{0}\neq b_{1}).\label{q1}
 \end{eqnarray}
Third, if Alice chooses $A_{3}$ and Bob chooses $B_{2}$, their measurement results are used to estimate the quantum phase-flip error rate $Q_{p1}$ as
\begin{eqnarray}
Q_{p1}=P(a_{3}\neq b_{2}).\label{q2}
 \end{eqnarray}
In the last case, if Alice chooses $A_{0}$ and Bob chooses $B_{2}$, or Alice chooses $A_{3}$ and Bob chooses $B_{1}$, their measurement results should be discarded.

 If $S_{1}\leq 2$ (the CHSH inequality), the measurement results from Alice and Bob  are classically correlated. Under this case, there exists a trivial attack for Eve to eavesdrop photons without being detected, so that the first photon transmission process is not secure and the parties have to discard the communication. If $S_{1}>2$, Alice's and Bob's measurement results are non-locally correlated, and they can bound Eve's photon interception rate. If $S_{1}$ reaches the maximal value of $2\sqrt{2}$, Alice and Bob share the maximally entangled state $|\phi^{+}\rangle_{AB}$. In this case, Eve cannot intercept any photon  without being detected.  As a result, when $2<S_{1}\leq 2\sqrt{2}$, Alice and Bob ensure that the first photon transmission process is secure and go on to the next step.

Step 3: Alice extracts the other stored photons in $S_{A1}$ sequence from the quantum memory and encodes her message on them by performing two unitary operations $U_{0}$ or $U_{1}$. The two unitary operations have the form of
 \begin{eqnarray}
U_{0}&=&\sigma_{x}=|V\rangle\langle H|+|H\rangle\langle V|,\nonumber\\
U_{1}&=&i\sigma_{y}=|H\rangle\langle V|-|V\rangle\langle H|,\label{encoding}
\end{eqnarray}
which can transform $|\phi^{+}\rangle$ to $|\psi^{+}\rangle$ and $|\psi^{-}\rangle$, respectively. Alice can encode her messages "0" and "1" on the photon pairs by performing $U_{0}$ and $U_{1}$, respectively. Meanwhile, Alice randomly selects some photons as the security checking photons for the second photon transmission round and does not perform any operation on them.
For preventing Eve to precisely intercept the corresponding encoded photons during the second photon transmission process according to her intercepted photons in the first photon transmission process, Alice messes up her photons in sequence $S_{A1}$ and records the position of each photon in the original sequence.

 Step 4: Alice successively sends all the photons in $S_{A1}$ sequence to Bob. After the photon transmission, Alice announces the position of each photon in the original $S_{A1}$ sequence by a public channel. Bob stores all the photons into the quantum memory devices and recovers the original photon sequence. Next, Alice announces the positions of the security checking photons and Bob extracts the security checking photon pairs in $S_{A1}$ and $S_{B1}$ sequences to make the second round of DI security checking by himself. After the measurements, Bob can estimate the CHSH polynomial $S_{2}$, the bit-flip error rate $Q_{b2}$, and the phase-flip error rate $Q_{p2}$. Similar as the  security checking, when $S_{2}\leq2$, the second photon transmission process is not secure and the parties should discard the communication. When $2<S_{2}\leq2\sqrt{2}$, they ensure that the second photon transmission process is secure and go on to the next step.

 Step 5: Bob extracts all the other photon pairs from the memory devices and makes the BSM on each of them. His BSM devices are the same as those in Charlie's location, which can only distinguish $|\psi^{+}\rangle$ and $|\psi^{-}\rangle$. After the measurement, Bob can read out the encoded messages by comparing his measurement results with the initial entangled state $|\phi^{+}\rangle$. For example, if the BSM result is $|\psi^{+}\rangle$, Bob can deduce that Alice performs $U_{0}$ operation, so that the encoded message is "0". If the BSM result is $|\psi^{-}\rangle$, Bob can obtain that Alice performs $U_{1}$ on the photon pair and the encoded message is "1".

\section{Security and communication quality of the DI-QSDC protocol against collective attacks}
In the device-independent scenario, Eve is only required to obey
the laws of quantum physics. In
both two security checking processes, Alice and Bob can only use
the observed correlations between the measurement basis
(input) and the measurement result (outcome) to bound Eve's knowledge. We consider a general attack, namely, collective attack, where Eve applies the same attack on each system of Alice
and Bob. As a result, after the photon transmission, all the
photon pairs have the same form. We also assume that each party's measurement result is only a
function of the current inputs.

We define that the secrete message capacity $C_{s}$ is the
amount of transmitted correct and secure qubits divided
by the total amount of the encoded photon pairs. Although we have specified a particular state
 in the DI-QSDC protocol to produce these correlations, we do not assume anything
about the implementation of the correlations when computing the secrete message capacity.

We first consider the ideal scenario, including the ideal devices and channels. If there is no eavesdropping, the CHSH polynomials in both security checking processes can reach the maximal value of $2\sqrt{2}$, and the bit-flip and phase-flip error rates are zero. In this case, any eavesdropping during the photon transmission processes would reduce the CHSH polynomials and increase the error rates, so that the eavesdropping can be easily detected. As a result, in the ideal scenario, Eve cannot eavesdrop any photon without being detected. As each encoded photon pair can transmit 1 bit of message, the value of $C_{s}$ equals to 1.

Next, we consider the practical scenario, including the practical devices and noisy channels. For collective attacks, the secrete message capacity from Alice to Bob is lower-bounded by the Devetak-Winter rate \cite{DIQKD2,DIQKD3} as
\begin{eqnarray}
C_{s}\geq I_{AB}-I_{AE},\label{Ec}
\end{eqnarray}
where $I_{AB}$ and $I_{AE}$ represent the mutual information between Alice and Bob, and the mutual information between Alice and Eve, respectively.
Since we have assumed uniform marginal, the mutual information between Alice and Bob
is given by \cite{DIQKD2,DIQKD3}
\begin{eqnarray}
I_{AB}=1-H(Q_{t}),
\end{eqnarray}
where $Q_{t}$ is the total error rate after two rounds of photon transmission, and $H(x)$ is the binary entropy with the form of
\begin{eqnarray}
H(x)=-x log_{2}x-(1-x) log_{2}(1-x).
\end{eqnarray}

In practical scenario, we have to consider the photon loss and decoherence. The photon loss can be divided into two categories, say, the transmission loss and local loss. The transmission loss represents the photon loss occurring in the transmission process. We provide the photon transmission efficiency $\eta_{t}=10^{-0.2L_{AB}/10}$ in Sec. II. The local loss represents all the photon loss occurring within the users' laboratories \cite{DIQKD10}. As the DI-QSDC requires the quantum memory devices, we have to consider the finite photon-extraction efficiency of the quantum memory. In this way, we define the local efficiency $\eta_{l}$ as the product of the coupling efficiency $\eta_{c}$ between the photon and the
fiber, the efficiency $\eta_{m}$ of the quantum memory, and the detection efficiency of the photon detector $\eta_{d}$ ($\eta_{l}=\eta_{c}\eta_{m}\eta_{d}$).
To our knowledge, the known DI protocols all require a high local efficiency, i.e. above 90\% \cite{DIQKD4,DIQKD7,DIQKD9}.
According to Sec. II, after the  round of photon transmission,  the photon transmission loss case can be eliminated with the help of the heralded architecture, but the local loss and decoherence still exist, which may degrade the entanglement and increase the total error rate. The decoherence caused by the channel noise has been widely researched in the quantum system \cite{decoherence}. Here, we consider a general model, say, the white noise model, in which the target state $|\phi^{+}\rangle$ may degrade to the other three Bell states in Eq. (\ref{bell}) with the same probability. After the first round of photon transmission, Alice and Bob can finally share $N_{1}$ pairs of mixed states as
 \begin{eqnarray}
\rho_{1}&=&\eta_{l}^{2}F|\phi^{+}\rangle\langle\phi^{+}|+\eta_{l}^{2}\frac{1-F}{3}(|\psi^{+}\rangle\langle\psi^{+}|+|\phi^{-}\rangle\langle\phi^{-}|\nonumber\\
&+&|\psi^{-}\rangle\langle\psi^{-}|)
+2\eta_{l}(1-\eta_{l})(|H\rangle\langle H|+|V\rangle\langle V|)\nonumber\\
&+&(1-\eta_{l})^{2}|vac\rangle\langle vac|,\label{rou1}
\end{eqnarray}
where $F$ is the fidelity of the target entangled state $|\phi^{+}\rangle$. If $|\phi^{+}\rangle$ transforms to the other state before the CBSM, Alice and Bob cannot deterministically construct the entanglement channel in $|\phi^{+}\rangle$ according to  the BSM result.
In theory, if there is no eavesdropping, the error rates and CHSH polynomial can be calculated as \cite{DIQKD2,DIQKD3},
\begin{eqnarray}
 Q_{b1}+Q_{p1}&=&\frac{1}{2}(1-\eta_{l}^{2})+\eta_{l}^{2}(1-F)\nonumber\\
 &=&\frac{1}{2}-\eta_{l}^{2}(F-\frac{1}{2}),\nonumber\\
 S_{1}&=&2\sqrt{2}\eta_{l}^{2}F.\label{S}
 \end{eqnarray}
During the second photon transmission process, the photon local loss, photon transmission loss and decoherence all may reduce the CHSH polynomial and increase the error rates. After the photon transmission, all the security checking
photon pairs have the form of
\begin{eqnarray}
\rho_{2}=\eta_{t}\eta_{l}^{2}\rho_{2}'+(1-\eta_{t}\eta_{l}^{2})\rho_{disturb},
 \end{eqnarray}
where
\begin{eqnarray}
\rho_{2}'&=&F^{2}|\phi^{+}\rangle\langle\phi^{+}|+\frac{1-F^{2}}{3}(|\psi^{+}\rangle\langle\psi^{+}|\nonumber\\
&+&|\phi^{-}\rangle\langle\phi^{-}|+|\psi^{-}\rangle\langle\psi^{-}|),\label{rou2}
\end{eqnarray}
and $\rho_{disturb}$ includes the disturb items, such as the single photon state and vacuum state. As a result, $S_{2}$ and $Q_{b2}+Q_{p2}$ can be written as
\begin{eqnarray}
Q_{b2}+Q_{p2}&=&\frac{1}{2}(1-\eta_{t}\eta_{l}^{2})+\eta_{t}\eta_{l}^{2}(1-F^{2})\nonumber\\
&=&\frac{1}{2}-\eta_{t}\eta_{l}^{2}(F^{2}-\frac{1}{2}),\nonumber\\
 S_{2}&=&2\sqrt{2}\eta_{t}\eta_{l}^{2}F^{2}.\label{Qt}
 \end{eqnarray}
After two rounds of photon transmission,  the total error rate $Q_{t}=Q_{b2}+Q_{p2}$. In this way, we can obtain
\begin{eqnarray}
I_{AB}=1-H(Q_{t})=1-H(Q_{b2}+Q_{p2}).
\end{eqnarray}

Then, we calculate $I_{AE}$. After the first and second photon transmission rounds, when $S_{1}>2$ and $S_{2}>2$, we can estimate the Holevo quantity  by
\begin{eqnarray}
\chi(S_{1})&\leq& H(\frac{1+\sqrt{(S_{1}/2)^{2}-1}}{2}),\nonumber\\
\chi(S_{2})&\leq& H(\frac{1+\sqrt{(S_{2}/2)^{2}-1}}{2}).\label{X}
\end{eqnarray}
The upper bound on the Holevo quantities in Eq. (\ref{X}) has been well proved and Eve's photon interception rates in the first and second photon transmission rounds equal to $\chi(S_{1})$ and $\chi(S_{2})$, respectively \cite{DIQKD2,DIQKD3}. It is obvious that $S_{2}<S_{1}$, so that we can obtain $\chi(S_{1})<\chi(S_{2})$. As Eve can read out the message only when she intercepts both photons of an encoded photon pair from Alice, we can bound the message leakage rate $I_{AE}$ of the current DI-QSDC protocol by
\begin{eqnarray}
I_{AE}\leq \chi(S_{1}).\label{IAE}
\end{eqnarray}
$I_{AE}$ reaches the maximum of $\chi(S_{1})$ only when in the second photon transmission round, Eve can intercept all the corresponding photons of her intercepted photons in the first photon transmission process. However, as Alice messes up her photons in sequence $S_{A1}$ before the second round of photon transmission, the probability that $I_{AE}$ reaches $\chi(S_{1})$ is quite close to 0 with a large number of transmitted photons.  Meanwhile, as $S_{1}$ is independent with the photon transmission efficiency $\eta_{t}$, the message leakage rate $I_{AE}$ is independent with the communication distance $L_{AB}$. \\

\begin{figure}[!h]
\begin{center}
\includegraphics[width=8cm,angle=0]{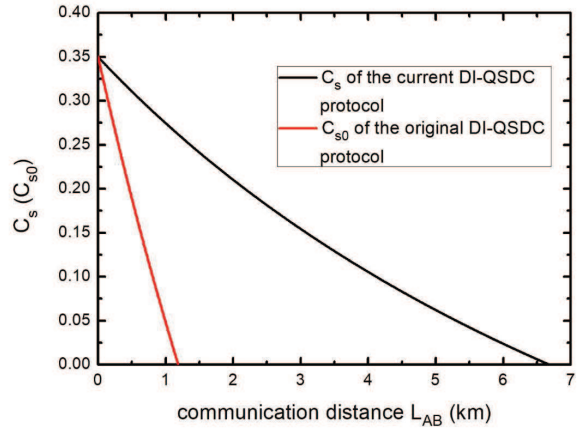}
\caption{The secrete message capacity $C_{s}$ of the current DI-QSDC protocol and $C_{s0}$ of the original DI-QSDC protocol in Ref. \cite{qsdc9}  as a function of the communication distance $L_{AB}$ in the device-independent scenario. Here, we control the local efficiency  $\eta_{l}=98\%$ and the fidelity $F=0.98$. }
\end{center}
\end{figure}

Based on above calculation, we provide the lower bound of $C_{s}$ in Eq. (\ref{Ec}) by
\begin{eqnarray}
C_{s}\geq 1-H(Q_{b2}+Q_{p2})-\chi(S_{1}).\label{Ecn}
\end{eqnarray}
As $Q_{t}$ increases with the growth of $L_{AB}$, $C_{s}$ would decrease with $L_{AB}$.
Different with quantum key distribution, as QSDC directly transmits secret messages, not the random keys, the parties cannot perform the post error correction method to correct the message error or message loss. We define the message loss rate ($r_{loss}$) as
the amount of lost message qubits divided by the total amount
of the message qubits, and the message error rate ($r_{error}$) as
the amount of incorrect qubits read out by Bob divided by the total
amount of the message qubits that Bob can read out. $r_{loss}$ and $r_{error}$ can be calculated as
\begin{eqnarray}
r_{loss}&=&1-\eta_{l}^{2}\eta_{t},\nonumber\\
r_{error}&=&1-F^{2}.
 \end{eqnarray}
As the photon transmission loss in the first photon transmission process does not cause message loss, the current DI-QSDC protocol has lower $r_{loss}$ than the original DI-QSDC protocol ($r_{loss0}=1-\eta_{l}^{2}\eta_{t}^{2}$) \cite{qsdc9}.

In Fig. 3, we provide  $C_{s}$ of the current DI-QSDC protocol and $C_{s0}$ of the original DI-QSDC protocol \cite{qsdc9} as a function of the communication distance $L_{AB}$ in the device-independent scenario. Here, we fix $F=0.98$ and $\eta_{l}=0.98$. It can be found that the maximal communication distance of the current DI-QSDC protocol can reach about 6.68 km, which is about 6 times of that in the original DI-QSDC protocol (about $1.18$ km). Meanwhile, at the same communication distance, $C_{s}$ is much higher than $C_{s0}$. \\

\begin{figure}[!h]
\begin{center}
\includegraphics[width=8cm,angle=0]{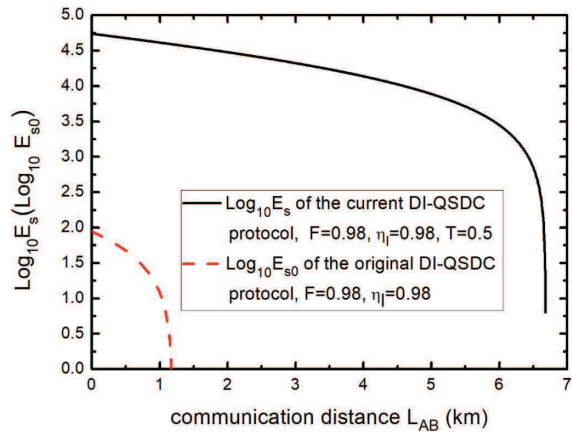}
\caption{The practical secure communication efficiency $E_{s}$ of the current DI-QSDC protocol and $E_{s0}$ of the original DI-QSDC protocol in Ref. \cite{qsdc9} on a logarithmic (with subscript 10)
versus the communication distance $L_{AB}$ in the device-independent scenario. Here, we control the local efficiency  $\eta_{l}=0.98$ and the fidelity $F=0.98$. We set the repetition rate $R_{rep}$ of both the single photon source and the spontaneous parametric down-conversion source to be 10 MHz. In current DI-QSDC protocol, we consider the transmittance of the VBS as $T=0.5$. In original DI-QSDC protocol, we consider the practical spontaneous parametric down-conversion source generates a pair of two-photon entangled state with the fidelity of $p=10^{-4}$.}
\end{center}
\end{figure}

Meanwhile, we define the practical secure communication efficiency $E_{s}$ as the
amount of transmitted correct and secure qubits per second. Here, we suppose the repetition rate of the photon source as $R_{rep}$. After each round of photon transmission, we choose half number of entangled photon pairs to make the DI security checking, so that only 1/4 amount of entangled photon pairs can be used to transmit messages. In this way, we can calculate $E_{s}$ of the current DI-QSDC protocol and $E_{s0}$ of the original DI-QSDC protocol as
\begin{eqnarray}
E_{s}&=&\frac{1}{4}R_{rep}P_{1}C_{s},\nonumber\\
E_{s0}&=&\frac{1}{4}R_{rep}C_{s0},
\end{eqnarray}
where $P_{1}$ is the success probability of the heralded BSM in Eq. (\ref{P1}). In Fig. 4, we show the $Log_{10}E_{s}$ of the current DI-QSDC protocol and $Log_{10}E_{s0}$ of the original DI-QSDC protocol \cite{qsdc9} as a function of  $L_{AB}$ in the device-independent scenario. Here, we set $F=0.98$ and $\eta_{l}=0.98$. The current DI-QSDC protocol adopts the on-demand single photon source and suitable VBSs with $T=0.5$. The original DI-QSDC protocol adopts the spontaneous parametric down-conversion source with the fidelity of $p=10^{-4}$. The repetition rate of both kinds of sources are set to be 10 MHz. It can be found that by adopting the single photon source and the heralded architecture, the practical secure communication efficiency of the current DI-QSDC protocol is about 600 times of that in the original DI-QSDC protocol.

\section{Discussion and conclusion}
 We propose a DI-QSDC protocol with single photon sources. In the protocol, the communication parties generate single photons from single photon sources and they can deterministically construct the entanglement channel with the help of the heralded BSM and the quantum memory. Then, Alice encodes the messages on her remained photons and sends the encoded photons to Bob for the BSM. Bob can finally read out the secret messages by comparing the BSM results with the original Bell state. By performing the DI security checking, the parties can guarantee the security of both photon transmission processes. Comparing with original DI-QSDC protocol \cite{qsdc9}, the current DI-QSDC protocol has two advantages. First, the practical single photon source is extremely close to the on-demand single photon source, and the adoption of the single photon source can increase the practical photon generation rate and eliminate security loophole from the double photon emission.
 Second, with the help of the heralded architecture, the parties can deterministically construct the entanglement channel from the single photons, and the message leakage rate of the DI-QSDC protocol is independent with the communication distance. Both  advantages enable the current DI-QSDC protocol to have much longer communication distance, much higher practical communication efficiency and lower message loss rate. It is noticeable that the DI-QKD protocol in Ref. \cite{DIQKD10} also adopts the BSM to heralded the construction of the entanglement channel. However, when two photons by one party are transmitted at the VBS and the other party has both photons reflected, the BSM may also obtain the successful measurement result but the parties cannot share the entanglement state. The probability that this happens scales exactly like $P_{1}$ even though $T\ll1$. In this way, the DI-QKD protocol in Ref. \cite{DIQKD10} cannot solve the double transmission interference problem, which may disturb the key generation. In our DI-QSDC protocol, the parties require to use quantum memory to store the reflected photons. Meanwhile, the quantum memory can also heralds the existence of the reflected photon. As a result, the double transmission cases can be eliminated, which is actually an attractive advantage of our DI-QSDC protocol comparing with the DI-QKD in Ref. \cite{DIQKD10}.

  As shown in Sec. IV, the decoherence occurring in both photon transmission processes may reduce $C_{s}$ and increase message error. Meanwhile, the decoherence also provides an opportunity for Eve to intercept some photons in both photon transmission processes without being detected. The entanglement purification is an effective method to resist the decoherence \cite{hyper6,EPP1,EPP2,EPP4}. In this way, we can adopt the entanglement purification in the current DI-QSDC protocol to improve the quality of entanglement channel. In detail, after Alice and Bob constructing the entanglement channel, they can perform the entanglement purification to increase the fidelity of $|\phi^{+}\rangle$. In theory, Alice and Bob can increase the fidelity of $|\phi^{+}\rangle$ to be quite close to 1 by repeating the entanglement purification. As a result, combined with the  entanglement purification, Alice and Bob can construct nearly perfect entanglement channel and obtain $S_{1}\rightarrow 2\sqrt{2}$. It makes Eve cannot intercept any photon during the first photon transmission process without being detected ($\chi(S_{1})\rightarrow 0$). As the message leakage rate $I_{AE}\leq\chi(S_{1})$, $I_{AE}$ can be reduced to 0 and the current DI-QSDC is absolutely secure. However, after the second transmission process, Bob cannot perform the entanglement purification, for the entanglement purification may change the encoded message. By performing the entanglement purification after the first photon transmission process, the total error rate of the current DI-QSDC protocol can be reduced from  $Q_{t}=\frac{1}{2}-\eta_{t}\eta_{l}^{2}(F^{2}-\frac{1}{2})$ to $Q_{t}'=\frac{1}{2}-\eta_{t}\eta_{l}^{2}(F-\frac{1}{2})$, and the lower
bound of the secrete message capacity can be increased to
\begin{eqnarray}
 C_{s}'\geq 1-H(Q_{t}').
 \end{eqnarray}

 Suppose Alice and Bob repeat the entanglement purification for $N$ times to construct the nearly perfect quantum channel, the practical communication efficiency of the current DI-QSDC protocol can be written as
 \begin{eqnarray}
 E_{sm}\geq \frac{1}{4}R_{rep}P_{1}\frac{P_{EP_{1}}P_{EP_{2}}\cdots P_{EP_{N}}}{2^{N}} [1-H(Q_{t}')],
 \end{eqnarray}
where $P_{EP_{i}}$ $(i=1,2,\cdots,N)$ represents the success probability of the ith round of entanglement purification.

In conclusion, DI-QSDC can resist all possible attacks on the imperfect experimental devices thus guarantee QSDC's security under practical imperfect experimental condition. The original DI-QSDC protocol adopts the practical entanglement photon source to generate the entanglement probabilistically, where the double-photon-pair emission is unavoidable. During the entanglement distribution process, the photon transmission loss largely deteriorates the entanglement. These two factors limit the secret message capacity and secure communication distance of the original DI-QSDC protocol. In the paper, we propose a DI-QSDC protocol with single photon sources and heralded architecture. The practical single photon source is extremely close to the on-demand single photon source. The parties can deterministically construct the entanglement channel from the single photons heralded by the BSM and the quantum memory. The security of the DI-QSDC protocol is guaranteed by the observation of data conclusively violating the CHSH inequality, so that it is unconditionally secure in theory. The photon transmission loss case in the first photon transmission process can be eliminated with the help of the heralded architecture, so that the message leakage rate is independent with the communication distance. Above two features can efficiently increase  DI-QSDC's secure communication distance and practical communication efficiency, and reduce the message loss rate. We numerically simulate the secrete message capacity and practical communication efficiency of our DI-QSDC protocol. Under the condition of $F=0.98$, $T=0.5$, and $\eta_{l}=0.98$, the maximal communication distance of the current DI-QSDC protocol reaches about 6.68 km, which is about 6 times of that in the original DI-QSDC protocol. The practical communication efficiency of the current DI-QSDC protocol is about 600 times of that in the original DI-QSDC protocol. Moreover, by performing the entanglement purification after the first photon transmission process, the parties can construct the nearly perfect entanglement channel, so that they can completely eliminate the message leakage and reduce the total error rate. Based on above features, this DI-QSDC protocol may have important application in future quantum communication field.

\section*{ACKNOWLEDGEMENTS} This work was supported by the National Natural Science Foundation
of China under Grant  Nos. 11974189 and 12175106.


\begin{thebibliography}{99}

\bibitem{BB84}  C. H. Bennett and G. Brassard, Quantum cryptography: public key distribution and coin tossing, Proceedings of the IEEE International Conference on Computers, Systems and Signal Processing, Bangalore, (India IEEE, New York, 1984), 175-179.
\bibitem{QKD1}  A. Ekert, Quantum cryptography based on Bell's theorem, Phys. Rev. Lett. \textbf{67}, 661 (1991).
\bibitem{addqkd4} P. W. Shor and J. Preskill, Simple proof of security of the BB84 quantum key distribution protocol, Phys. Rev. Lett. \textbf{85}, 441-444 (2000).
\bibitem{QKD2}  R. Ursin, F. Tiefenbacher, T. Schmitt-Manderbach, H. Weier, T. Scheidl, M. Lindenthal, B. Blauensteiner, T. Jennewein, J. Perdigues, P. Trojek, B. Omer, M. Furst, M. Meyenburg, J. Rarity, Z. Sodnik, C. Barbieri, H. Weinfurter, and A. Zeilinger, Entanglement-based quantum communication over 144 km, Nat. Phys. \textbf{3}, 481 (2007).
\bibitem{QKD3}  T. Sasaki, Y. Yamamoto, and M. Koashi, Practical quantum key distribution protocol without monitoring signal disturbance, Nature \textbf{509}, 475-478 (2014).
\bibitem{QKD4}  S. Wang, Z. Q. Yin, W. Chen, D. Y. He, X. T. Song, H. W. Li, L. J. Zhang, Z. Zhou, G. C. Guo, and Z. F. Han, Experimental demonstration of a quantum key distribution without signal disturbance monitoring, Nat. Photon. \textbf{9}, 832-836 (2015).

\bibitem{QKD7} F. H. Xu, X. F. Ma, Q. Zhang, H. K. Lo, and J. W. Pan, Secure quantum key distribution with realistic devices, Rev. Mod. Phys. \textbf{92}, 025002 (2020).

\bibitem{qkd7nn} S. Zhao, P. Zeng, W. F. Cao, X. Y. Xu, Y. Z. Zhen, X. F. Ma, L. Li, N. L. Liu, and K. Chen, Phase-matching quantum cryptographic conferencing, Phys. Rev. Appl. \textbf{14}, 024010 (2020).

\bibitem{qkd7n} A. R. Jin, P. Zeng, R. V. Penty, and X. F. Ma, Reference-frame-independent design of phase-matching quantum key distribution, Phys. Rev. Appl. \textbf{16}, 034017 (2021).

\bibitem{QKD8} Y. A. Chen, Q. Zhang, T. Y. Chen, W. Q. Cai, S. K. Liao, J. Zhang, K. Chen, J. Yin, J. G. Ren, Z. Chen, \emph{et al.,} An integrated space-to-ground quantum communication network over 4,600 kilometres, Nature  \textbf{589}, 214-219 (2021).

\bibitem{QKD17n} W. B. Liu, C. L. Li, Y. M. Xie, C. X. Weng, J. Gu, X. Y. Cao, Y. S. Lu, B. H. Li, H. L. Yin, and Z. B. Chen, Homodyne detection quadrature phase shift keying continuous-variable quantum key distribution with high excess noise tolerance, PRX Quant. \textbf{2}, 040334 (2021).

\bibitem{qkd9} Z. Q. Yin, F. Y. Lu, J. Teng, S. Wang, W. Chen, G. C. Guo, and Z. F. Han, Twin-field protocols: towards intercity quantum key distribution without quantum repeaters, Funda. Res. \textbf{1}, 93-95 (2021).
		
\bibitem{qkd10} L. C. Kwek, L. Cao, W. Luo, Y. X. Wang, S. H. Sun, X. B. Wang, and A. Q. Liu, Chip-based quantum key distribution, AAPPS Bull. \textbf{31}, 15 (2021).

\bibitem{qkd11} H. Guo, Z. Y. Li, S. Yu, and Y. C. Zhang, Toward practical quantum key distribution using telecom components, Founda. Res. \textbf{1}, 96-98 (2021).

\bibitem{qkd12} G. Z. Tang, C. Y. Li, and M. Wang, Polarization discriminated time-bin phase-encoding measurement-device-independent quantum key distribution, Quant. Eng. \textbf{3}, e79 (2021).
\bibitem{qkd13} X. F. Wang, X. J. Sun, Y. X. Liu,  W. Wang, B. X. Kan, P. Dong, and L. L. Zhao, Transmission of photonic polarization states from geosynchronous earth orbit satellite to the ground, Quant. Eng. \textbf{3}, e73 (2021).

\bibitem{qkd14} C. Y. Zhang and Z. J. Zheng, Entanglement-based quantum key distribution with untrusted third party. Quant. Inform. Process. \textbf{20}, 146 (2021).

\bibitem{qkd15} W. Zhao, R. H. Shi, X. C. Ruan, Y. Guo, Y. Y. Mao, and Y. Y. Feng, Monte Carlo-based security analysis for multi-mode continuous-variable quantum key distribution over underwater channel, Quant. Inform. Process. \textbf{21}, 186 (2022).

\bibitem{qkd14n} C. Zhou, X. Y. Wang, Z. G. Zhang, S. Yu, Z. Y. Chen, and H. Guo, Rate compatible reconciliation for continuous-variable quantum key distribution using raptor-like LDPC codes,  Sci. China Phys. Mech. \& Astron. \textbf{64}, 260311 (2022).

\bibitem{qkd15n}  B. Liu, S. Xia, D. Xiao, W. Huang, B. J. Xu, and Y. Li, Decoy-state method for quantum-key-distribution-based quantum private query, Sci. China Phys. Mech. \& Astron. \textbf{65}, 240312 (2022).

\bibitem{QKD15n} Y. M. Xie, Y. S. Lu, C. X. Weng, X. Y. Cao, Z. Y. Jia, Y. Bao, Y. Wang, Y. Fu, H. L. Yin, and Z. B. Chen, Breaking the rate-loss bound of quantum key distribution with asynchronous two-photon interference, PRX Quant. \textbf{3}, 020315 (2022).

\bibitem{qsdc1} G. L. Long and X. S. Liu, Theoretically efficient high-capacity quantum-key-distribution scheme, arXiv: preprint quant-ph/0012056, (2000) (Phys. Rev. A \textbf{65}, 032302 (2002)).

\bibitem{qsdc2} F. G. Deng, G. L. Long, and X. S. Liu, Two-step quantum direct communication protocol using the Einstein-Podolsky-Rosen pair block, Phys. Rev. A \textbf{68}, 042317 (2003).

\bibitem{qsdc3} F. G. Deng and G. L. Long, Secure direct communication with a quantum one-time pad, Phys. Rev. A \textbf{69}, 052319 (2004).

\bibitem{qsdc4n} C. Wang, F. G. Deng, Y. S. Li, X. S. Liu, and G. L. Long, Quantum secure direct communication with high-dimension quantum superdense coding, Phys. Rev. A \textbf{71}, 044305 (2005).

\bibitem{qsdc4} J. Y. Hu, B. Yu, M. Y. Jing, L. T. Xiao, S. T. Jia, G. Q. Qin, and G. L. Long, Experimental quantum secure direct communication with single photons, Light Sci. Appl. \textbf{5}, e16144 (2016).

\bibitem{qsdc5} W. Zhang, D. S. Ding,  Y. B. Sheng,  L. Zhou, B. S. Shi, and G. C. Guo, Quantum secure direct communication with quantum memory, Phys. Rev. Lett. \textbf{118}, 220501 (2017).
\bibitem{qsdc6} F. Zhu, W. Zhang, Y. B. Sheng, and Y. D. Huang, Experimental long-distance quantum secure direct communication, Sci. Bull. \textbf{62}, 1519-1524 (2017).

\bibitem{qsdc8} R. Y. Qi, Z. Sun, Z. S. Lin, P. H. Niu, W. T. Hao, L. Y. Song, Q. Huang, J. C. Gao, L. G. Yin, and G. L. Long, Implementation and security analysis of practical quantum secure direct communication, Light Sci. Appl. \textbf{8}, 22 (2019).
\bibitem{qsdc9} L. Zhou, Y. B. Sheng, and G. L. Long, Device-independent quantum secure direct communication against collective attacks, Sci. Bull. \textbf{65}, 12-20 (2020).
\bibitem{qsdc10} Z. R. Zhou, Y. B. Sheng, P. H. Niu,  L. G. Yin, G. L. Long, and L. Hanzo, Measurement-device-independent quantum secure direct communication, Sci. China Phys. Mech. \& Astron. \textbf{63}, 230362 (2020).
\bibitem{addqsdc1} Z. Sun, L. Y. Song, Q. Huang, L. G. Yin, G. L. Long, J. H. Lu, and L. Hanzo, Toward practical quantum secure direct communication: a quantum-momery-free protocol and code design, IEEE Tran. Commun. \textbf{68}, 5778-5792 (2020).

\bibitem{addqsdc2} D. Pan, Z. S. Lin, J. W. Wu, H. R. Zhang, Z. Sun, D. Ruan, L. G. Yin, and G. L. Long,  Experimental free-space quantum secure direct communication and its security analysis, Photon. Res. \textbf{8}, 1522-1531 (2020).

\bibitem{addqsdc3} L. Yang, J. W. Wu, Z. S. Lin, L. G. Yin, and G. L. Long, Quantum secure direct communication with entanglement source and single-photon measurement, Sci. China Phys. Mech. \& Astron. \textbf{63}, 110311 (2020).

\bibitem{qsdc11} T. Li and G. L. Long, Quantum secure direct communication based on single-photon bell-state measurement, New J. Phys. \textbf{22}, 063017 (2020).

\bibitem{qsdc12} G. L. Long and H. R. Zhang, Drastic increase of channel capacity in quantum secure direct communication using masking, Sci. Bull. \textbf{66}, 1267-1269 (2021).

\bibitem{qsdc12n} X. Liu, Z. J. Li, D. Luo, C. F. Huang, D. Ma, M. M. Geng, J. W. Wang, Z. R. Zhang,  and K. J. Wei, Practical decoy-state quantum secure direct communication, Sci. China Phys. Mech. \& Astron. \textbf{64}, 120311 (2021).

\bibitem{qsdc13} Z. W. Cao, L. Wang, K. X. Liang, G. Chai, and J. Y. Peng, Continuous-variable quantum secure direct communication based on Gaussian mapping, Phys. Rev. Appl. \textbf{16}, 024012 (2021).

\bibitem{qsdc14} Z. T. Qi, Y. H. Li, Y. W. Huang, J. Feng, Y. L. Zheng, and X. F. Chen, A 15-user quantum secure direct communication network, Light Sci. Appl. \textbf{10}, 183 (2021).

\bibitem{qsdc14n} Z. M. Huang, Z. B. Rong, X. F. Zou, and Z. M. He, Semi-quantum secure direct communication in the curved spacetime, Quant. Inform. Process. \textbf{20}, 375 (2021).

\bibitem{qsdc15} H. R. Zhang, Z. Sun, R. Y. Qi,  L. G. Yin, G. L. Long, and J. H. Lu, Realization of quantum secure direct communication over 100 km fiber with time-bin and phase quantum states, Light Sci. Appl. \textbf{11}, 83 (2022).


\bibitem{qsdc16n} L. Liu, B. Lu, J. Y. Song, and C. Wang, Secure communications based on sending-or-not-sending strategy, Quant. Inform. Process. \textbf{21}, 250 (2022).

\bibitem{qsdc15n} N. Das and G. Paul, Measurement device-independent quantum secure direct communication with user authentication, Quant. Inform. Process. \textbf{21}, 260 (2022).

\bibitem{qsdc16} Y. B. Sheng, L. Zhou, and G. L. Long, One-step quantum secure direct communication, Sci. Bull. \textbf{67}, 367-374 (2022).

\bibitem{qsdc17} L. Zhou and Y. B. Sheng, One-step device-independent quantum secure direct communication, Sci. China Phys. Mech. \& Astron. \textbf{65}, 250311 (2022).

\bibitem{qsdc18} J. W. Wu, G. L. Long, and M. Hayashi, Quantum secure direct communication with private dense coding using a general preshared quantum state, Phys. Rev. Appl. \textbf{17}, 064011 (2022).

\bibitem{qsdc19} G. L. Long, D. Pan, Y. B. Sheng, Q. K. Xue, J. H. Lu, and L. Hanzo, An evolutionary pathway for the quantum internet relying on secure classical
repeaters, IEEE Network \textbf{36}, 82-88 (2022).

\bibitem{DIQKD2} A. Ac\'{i}n, N. Brunner, N. Gisin, S. Massar, S. Pironio, and V. Scarani, Device-independent security of quantum cryptography against collective attacks, Phys. Rev. Lett. \textbf{98}, 230501 (2007).
\bibitem{DIQKD3} S. Pironio, A. Ac\'{i}n, N. Brunner, N. Gisin, S. Massar, and V. Scarani, Device-independent quantum key distribution secure against collective attacks, New J. Phys. \textbf{11}, 045021 (2009).
\bibitem{DIQKD4} N. Gisin, S. Pironio, and N. Sangouard, Proposal for implementing device-independent quantum key distribution based on a heralded qubit amplifier, Phys. Rev. Lett.  \textbf{105}, 070501 (2010).
\bibitem{DIQKD6} C. C. W. Lim, C. Portmann, M. Tomamichel, R. Renner, and N. Gisin, Device-independent quantum key distribution with local Bell test, Phys. Rev. X, \textbf{3}, 031006 (2013).
\bibitem{DIQKD7} K. P. Seshadreesan, M. Takeoka, and M. Sasaki, Progress towards practical device-independent quantum key distribution with spontaneous parametric down-conversion sources, on-off photodetectors, and entanglement swapping, Phys. Rev. A, \textbf{93}, 042328 (2016).
\bibitem{DIQKD8} R. Arnon-Friedman, F. Dupuis, O. Fawzi, R. Renner, and T. Vidick, Practical device-independent quantum cryptography via entropy accumulation, Nat. Commun. \textbf{9}, 459 (2018).

\bibitem{DIQKD9} V. Zapatero and M. Curty, Long-distance device-independent quantum key distribution, Sci. Rep. \textbf{9}, 17749 (2019).

\bibitem{DIQKD10} J. Ko{\l}ody\'{n}ski, A. M\'{a}ttar, P. Skrzypczyk, E. Woodhead, D. Cavalcanti, K. Banaszek, and A. Ac\'{i}n, Device-independent quantum key distribution with single-photon sources, Quantum \textbf{4}, 260 (2020).

\bibitem{Bell1}  J. S. Bell, On the Einstein-Podolsky-Rosen paradox, Physics \textbf{1}, 195-200 (1964).

\bibitem{Bell2} N. Brunner, D. Cavalcanti, S. Pironio, V. Scarani, and S. Wehner, Bell nonlocality, Rev. Mod. Phys. \textbf{86}, 419 (2014).

\bibitem{CHSH1}  J. F. Clauser, M. A. Horne, A. Shimony, and R. A. Holt, Proposed experiment to test local hidden-variable theories, Phys. Rev. Lett. \textbf{23}, 880 (1969).

\bibitem{detector1} W. J. Zhang, L. X. You, H. Li, J. Huang, C. L. Lv, L. Zhang, X. Y. Liu,
J. J. Wu, Z. Wang, and X. M. Xie, NbN superconducting nanowire single photon detector with efficiency over 90\% at 1550 nm wavelength operational at compact cryocooler temperature, Sci. China Phys. Mech. \& Astron. \textbf{60}, 120314 (2017).

\bibitem{detector2} X. Y. Lu, Q. Li, D. A. Westly, G. Moille, A. Singh, V. Anant, and K. Srinivasan, Chip-integrated visible-telecom entangled photon pair source for quantum communication, Nat. Phys. \textbf{15}, 373-381 (2019).

\bibitem{spdc} P. G. Kwiat, K. Mattle, H. Weinfurter, and A. Zeilinger, New high-intensity source of polarization-entangled photon pairs,  Phys. Rev. Lett. \textbf{116}, 213601 (2016).

\bibitem{spdc1} L. K. Chen, H. L. Yong, P. Xu, X. C. Yao, T. Xiang, Z. D. Li, C. Liu, H. Lu, N. L. Liu, L. Li, T. Yang, C. Z. Peng, B. Zhao, Y. A. Chen, and J. W. Pan, Experimental nested purification for a linear optical quantum repeater, Nat. Photon. \textbf{11}, 695-699 (2017).
\bibitem{hyper6} X. M. Hu, C. X. Huang, Y. B. Sheng, L. Zhou, B. H. Liu, Y. Guo, C. Zhang, W. B. Xing, Y. F. Huang, C. F. Li, and G. C. Guo, Long-distance entanglement purification for quantum communication, Phys. Rev. Lett. \textbf{126}, 010503 (2021).


\bibitem{singlephoton1} M. M\"{u}ller, S. Bounouar, K. D. J\"{o}ns, M. Gl\"{a}ssl, and P. Michler, On-demand generation of indistinguishable polarization-entangled photon
pairs, Nat. Photon. \textbf{8}, 224-228 (2014).

\bibitem{singlephoton2} J. Claudon, J. Bleuse, N. S. Malik, M. Bazin, P. Jaffrennou, N. Gregersen, C. Sauvan, P. Lalanne, and J. M. Gerard, A highly efficient
single-photon source based on a quantum dot in a photonic nanowire, Nat. Photon.
\textbf{4}, 174-177 (2010).

\bibitem{singlephoton3} J. C. Loredo, N. A. Zakaria, N. Somaschi, C. Anton, L. de Santis, V. Giesz, T. Grange, M. A. Broome, O. Gazzano, G. Coppola, I. Sagnes, A. Lemaitre, A. Auffeves, P. Senellart, 
M. P. Almeida, and A. G. White, Scalable performance in solid-state single-photon sources, Optica \textbf{3}, 433-440 (2016).

\bibitem{singlephoton4} H. Wang, Z. C. Duan, Y. H. Li, S. Chen, J. P. Li, Y. M. He, M. C. Chen, Y. He, X. Ding, C. Z. Peng, C. Schneider, M. Kamp, S. H\"{o}fling, C. Y. Lu, and J. W. Pan, Near-transform-limited single photons from an efficient solid-state quantum emitter, Phys. Rev. Lett. \textbf{116}, 213601 (2016).

\bibitem{singlephoton5} J. H. Kim, T. Cai, C. J. K. Richardson, R. P. Leavitt, and E. Waks, Two-photon interference from a bright single-photon source at telecom wavelengths, Optica \textbf{3}, 577-584 (2016).

\bibitem{singlephoton6} N. Somaschi, V. Giesz, L. De Santis, J. C. Loredo, M. P. Almeida, G. Hornecker, S. L. Portalupi, T. Grange, C. Anton, J. Demory,
C. Gomez, I. Sagnes, N. D. Lanzillotti-Kimura, A. Lemaitre, A. Auffeves, A. G. White, L. Lanco, and P. Senellart, Near-optimal single-photon sources in the solid state, Nat. Photon. \textbf{10}, 340-345 (2016).

\bibitem{singlephoton7} X. Ding, Y. He, Z. C. Duan, N. Gregersen, M. C. Chen, S. Unsleber, S. Maier, C. Schneider, M. Kamp, S. H\"{o}fling, C. Y. Lu, and J. W. Pan, On-demand single
photons with high extraction efficiency and near-unity indistinguishability from a resonantly driven quantum dot in a micropillar, Phys. Rev. Lett. \textbf{116}, 020401 (2016).

\bibitem{source7} S. Shi, B. Xu, K. Zhang, G. S. Ye, D. S. Xiang, Y. B. Liu, J. Z. Wang, D. Q. Su, and L. Li, High-fidelity photonic quantum logic gate
based on near-optimal Rydberg single-photon source, Nat. Commun. \textbf{13}, 4454 (2022).

\bibitem{source8} A. Barbiero, J. Huwer, J. Skiba-Szymanska, D. J. P. Ellis, R. M. Stevenson,
T. M\"{u}ller, G. Shooter, L. E. Goff, D. A. Ritchie, and A. J. Shields, High-performance single-photon sources at telecom wavelength
based on broadband hybrid circular bragg gratings. ACS Photon. \textbf{9}, 3060-3066 (2022).

\bibitem{entanglementpra} M. Lasota, C. Radzewicz, K. Banaszek, and R. Thew, Linear optics schemes for entanglement distribution with realistic single-photon sources, Phys. Rev. A \textbf{90},
033836 (2014).

\bibitem{memory} D. S. Ding, Z. Y. Zhou, B. S. Shi, and G. C. Guo, Single-photon-level quantum image memory based on cold atomic ensembles, Nat. Commun. \textbf{4}, 2527 (2013).

\bibitem{memory1} E. Distante, P. Farrera, A. Padr\'{}n-Brito, D. Paredes-Barato, G. Heinze, H. de Riedmatten, Storing single photons emitted by a quantum memory on a highly excited Rydberg state, Nat. Commun. \textbf{8}, 14072 (2017).

\bibitem{loss} V. Scarani, H. Bechmann-Pasquinucci, N. J. Cerf, M. Dusek, N. Lutkenhaus, and M. Peev, The security of practical quantum key distribution, Rev. Mod. Phys. \textbf{81}, 1301 (2009).

\bibitem{decoherence} B. X. Wang, M. J. Tao, Q. Ai, T. Xin, N. Lambert, D. Ruan, Y. C. Cheng, F. Nori, F. G. Deng, and G. L. Long, Efficient quantum simulation of photosynthetic light harvesting, npj Quant. Inf. \textbf{4}, 52 (2018).

\bibitem{EPP1} C. H. Bennett, G. Brassard, S. Popescu, B. Schumacher, J. A. Smolin, and W. K. Wootters, Purification of noisy entanglement and faithful teleportation via noisy channels, Phys. Rev. Lett. \textbf{76}, 722 (1996).

\bibitem{EPP2} J. W. Pan, C. Simon, C. Brukner, and A. Zeilinger, Entanglement purification for quantum communication, Nature \textbf{410}, 1067-1070 (2001).

\bibitem{EPP4} C. X. Huang, X. M. Hu, B. H. Liu, L. Zhou, Y. B. Sheng, C. F. Li, and G. C. Guo, Experimental one-step deterministic polarization entanglement purification, Sci. Bull. \textbf{67},
593 (2022).

\end{thebibliography}
\end{document}